\documentclass[conference]{IEEEtran}
\IEEEoverridecommandlockouts     % onecolumn (second format)

\usepackage{subcaption}
\usepackage{verbatim}
\usepackage{caption}
\usepackage{subcaption}
\usepackage[]{algorithm2e}

\usepackage{booktabs}
\usepackage{relsize}
\usepackage{colortbl}
\usepackage{enumerate}
\usepackage{amsmath}
\usepackage{tikz}
\usepackage{pgf}
\usepackage{pbox}
\usepackage{rotating}
\usepackage{multirow}
\usepackage{balance}
\usepackage{tablefootnote}
\usepackage[bookmarks=false]{hyperref}
\usepackage{flushend}
\usepackage{tcolorbox}
\usepackage[justification=centering]{caption}% or e.g. [format=hang]

\newcommand{\smallsection}[1]{\noindent \textbf{\underline{#1}}.}
\newcommand{\ea}{{\em et al.}}
\newcommand{\finding}[1]{
\begin{tcolorbox}[leftrule=1mm,toprule=0mm,bottomrule=0mm,left=1pt,right=2pt,top=2pt,bottom=2pt]
\em #1
\end{tcolorbox}
}

\newcommand{\rqone}{Is the attention mechanism suitable to explain the pre-trained code models?}
\newcommand{\rqtwo}{What do the pre-trained code models learn?}
\newcommand{\rqthree}{When do the pre-trained code models not work?}

\usepackage{amsmath,amssymb,amsfonts}
\usepackage{algorithmic}
\usepackage{graphicx}
\usepackage{textcomp}
\usepackage{xcolor}

\usepackage{listings}
\usepackage{color}
\definecolor{dkgreen}{rgb}{0,0.6,0}
\definecolor{gray}{rgb}{0.5,0.5,0.5}
\definecolor{mauve}{rgb}{0.58,0,0.82}
\usepackage{caption}
\definecolor{codesColor}{rgb}{0.95, 0.95, 0.95}
\begin{document}

\title{Explaining Transformer-based Code Models: \\ What Do They Learn? When They Do Not Work?}

\author{\IEEEauthorblockN{1\textsuperscript{st} Ahmad Haji Mohammadkhani}
\IEEEauthorblockA{\textit{University of Calgary} \\
Calgary, Canada \\
ahmad.hajimohammadkh@ucalgary.ca}
\and
\IEEEauthorblockN{2\textsuperscript{nd} Chakkrit Tantithamthavorn}
\IEEEauthorblockA{\textit{Monash University} \\
Melbourne, Australia \\
Chakkrit@monash.edu}
\and
\IEEEauthorblockN{3\textsuperscript{rd} Hadi Hemmatif}
\IEEEauthorblockA{\textit{York University} \\
Toronto, Canada \\
hemmati@yorku.ca}
}
\maketitle

\begin{abstract}
In recent years, there has been a wide interest in designing deep neural network-based models that automate downstream software engineering tasks on source code, such as code document generation, code search, and program repair. Although the main objective of these studies is to improve the effectiveness of the downstream task, many studies only attempt to employ the next best neural network model, without a proper in-depth analysis of why a particular solution works or does not, on particular tasks or scenarios. In this paper, using an example eXplainable AI (XAI) method (attention mechanism), we study two recent large language models (LLMs) for code (CodeBERT and GraphCodeBERT) on a set of software engineering downstream tasks: code document generation (CDG), code refinement (CR), and code translation (CT). Through quantitative and qualitative studies, we identify what CodeBERT and GraphCodeBERT learn (put the highest attention on, in terms of source code token types), on these tasks. We also show some of the common patterns when the model does not work as expected (performs poorly even on easy problems) and suggest recommendations that may alleviate the observed challenges. 
\end{abstract}

\begin{IEEEkeywords}
Explainable AI (XAI), LLM, Code Models, Interpretable, Attention, Transformer.
\end{IEEEkeywords}

\section{Introduction}

Large Language Models (LLMs) for code (in short: code models) are proposed to analyze the big corpora of source code and natural languages collected from open-source platforms (e.g., GitHub and StackOverflow).
Such pre-trained code models have been used to automate various source code-related tasks, e.g., code understanding, code generation, code clone detection ~\cite{shobha2021code}, defect detection ~\cite{pornprasit2021pyexplainer,humphreys2019explainable}, and code summarization ~\cite{leclair2020improved}.
Automating such software engineering tasks has been shown to greatly improve software developers' productivity and reduce the costs of software development.

Recent studies proposed Transformer-based code models, e.g., CodeBERT ~\cite{feng2020codebert}, GraphCodeBERT ~\cite{guo2020graphcodebert}, CodeGPT ~\cite{lu2021codexglue}, CodeT5 ~\cite{wang2021codet5}.
However, most of these studies often focus on improving its accuracy---\emph{\textbf{without considering its explainability aspect}}.
Thus, when deploying such models in practice, practitioners still do not know why such models provide a given recommendation or suggestion.

Let's consider a given Python code snippet of a bubble sort algorithm.
A code summarization model may be able to correctly summarize that the given code snippet is a bubble sort algorithm.
However, developers may not trust the models if the models correctly generate the natural text based on indentation, white spaces, or parenthesis of the Python code snippet, instead of the meaningful semantic information (i.e., bubble sort).
Thus, the correct predictions generated by the models do not guarantee that the models are learned correctly.
Therefore, a lack of explainability of the large and complex code models could lead to a lack of adoption in practice.

\emph{In this paper}, we conduct an empirical study to analyze code models through the lens of Explainable AI. 
Particularly, we focus on the two well-known code models, i.e., CodeBERT and GraphCodeBERT with the three understanding \& generation-specific downstream tasks, i.e., Code Summarization (Code$\rightarrow$Text), Code Transformation (Code$\rightarrow$Code), and Code Translation (Code$\rightarrow$Code) tasks.
To explain the predictions of these models, we leverage an attention mechanism inside the Transformer architecture, which is an intrinsic Explainable AI approach. 
The attention mechanism allows us to understand what are the most important tokens in the input sequence that contribute the most to the tokens in the output sequences.
In particular, we aim to address the following two research questions:

%\begin{enumerate}%[{\bf(RQ1)}]

%\item {\bf \rqone}\\
%\smallsection{Results} 
%The results show that in a notable proportion of the outputs generated by the models, the chosen token already exists in the input and it has a high attention score. As the average ``normalized attention rank'' of the chosen token in the last layers of the model is less than 3\%, 4\%, and 13\% for code translation, code refinement, and code document generation, respectively. We conclude that the attention mechanism has a considerable weight in the model's decision making and hence, it can be a good suitable tool to explain it.

%\item {}\\
{\bf \rqtwo}\\
%\smallsection
{\bf Results:} 
Analyzing attention scores and their distribution over different token types shows that the models learn to focus on specific types of tokens for each downstream task. In CDG, the models learn to focus on the methods signatures (i.e., method name and input arguments). While in CT, syntax, that is tokens related to the programming language, attract more attention. CR has a middle ground compared to the other two tasks and has a more balanced distribution of attention. Also, it is shown that GraphCodeBERT pays more attention to the structural parts of the source code, rather than CodeBERT, which likely is the result of the additional step of GraphCodeBERT for parsing the code and leveraging the code's data flow. 
These observations are inline with what is expected from a code model which led us to conclude that the two studied models are indeed learning (in most cases) what they are supposed to.

%\item {}\\
{\bf \rqthree}\\
\smallsection{Results} 
Our finding shows that there are certain situations that cause the models to perform poorly across different tasks. For instance, the models do not work well with samples having long or complex source code and/or long expected answers (model outputs). We also showed that poor performance of the model is usually projected in its attention distribution. In other words, whenever the model fails to achieve a good output for a model, it also has failed to pay enough attention to the corresponding token types for the respective downstream task. 
We have also provided some recommendations on how to potentially alleviate these weaknesses in the paper.

% \end{enumerate}

These findings lead us to conclude that even though pre-trained models have shown great results on software engineering tasks; none of them can be considered a closed problem and there are certain aspects of these models that need more focus through further studies. Explaining these models can shed light on their weaknesses and provide directions for future research.

% \smallsection{Contributions} Contributions of this paper are as follows:

% \begin{itemize}
    % \item We empirically study the suitability of the attention mechanism as an XAI method to explain Transformer-based models.
%     \item Analyzing the attention scores, we found interesting insights into models' decision-making for different tasks.
%     \item We provide explanations on several scenarios, where CodeBERT and GraphCodeBERT under-perform.
%     \item We offer some actionable recommendations on how to improve the models, in the future, to potentially alleviate the observed weaknesses.
% \end{itemize}    

\smallsection{Open Science} To foster the open science initiative, we made the replication package publicly available at GitHub.\footnote{https://anonymous.4open.science/r/XAI-for-transformer-models-F26A}
% \footnote{https://github.com/Ahmad535353/XAI-for-transformer-models}

% \smallsection{Paper Organization} 
% In the rest of this paper, in Section~\ref{background}, we briefly introduce the background and related work on Transformer-based code models and XAI in software engineering. In Section~\ref{Motivating}, we provide motivating examples about some observed weaknesses in code models that could use some explanation. Section~\ref{research_method} explains our research method by going through the objectives and design details and finally Section~\ref{results} demonstrates and discusses the results of our experiments and Section~\ref{limitations} explains the possible limitations of this work that may get improved in the future works and finally Section~\ref{Conclusion} concludes the paper. 

\section{Background \& Related Work} \label{background}

%In this section, we present background knowledge and related work on Explainable AI in software engineering.

Explainability is now becoming a critical concern in software engineering.
Many researchers often employed AI/ML techniques for defect prediction, malware detection, and effort estimation.
While these AI/ML techniques can greatly improve developers' productivity, software quality, and end-user experience, practitioners still do not understand why such AI/ML models made those predictions ~\cite{tantithamthavorn2021actionable,tantithamthavorn2021explainable,jiarpakdee2021practitioners,jiarpakdee2020empirical,rajapaksha2021sqaplanner,pornprasit2021pyexplainer}.
To address this challenge, researchers propose various approaches to generate explanations at two levels:

\emph{(1) Global explanations} can be generated using interpretable machine learning techniques (e.g., decision tree, decision rules, and logistic regression techniques) or intrinsic model-specific techniques (e.g., ANOVA, variable importance) so the entire predictions and recommendations process is transparent and comprehensible.
However, such intrinsic model-specific techniques aim to provide global explainability, without providing explanations to individual predictions.

\emph{(2) Local explanations}, on the other hand, can be generated using some techniques (e.g., LIME, SHAP) to explain the predictions of complex black-box AI/ML models (e.g., neural network, random forest).
Such techniques can provide an explanation for each individual prediction (i.e., an instance to be explained), allowing users to better understand why the prediction is made.% by the AI/ML models.

In software engineering, explainable AI has been recently studied in the domain of defect prediction (i.e., a classification model to predict if a file/class/method will be defective in the future or not).
In particular, the survey study by Jiarpakdee~\ea ~\cite{jiarpakdee2021practitioners} found that explaining the predictions is as equally important and useful as improving the accuracy of defect prediction.
However, their literature review found that 91\% (81/96) of the defect prediction studies only focus on improving the predictive accuracy, without considering explaining the predictions, while only 4\% of these 96 studies focus on explaining the predictions.

Although XAI is still a very under-researched topic within the software engineering community, very few existing XAI studies have shown some successful usages e.g., in defect prediction. In one example, Wattanakriengkrai~\ea ~\cite{wattanakriengkrai2020predicting} and Pornprasit and Tantithamthavorn ~\cite{pornprasit2021jitline} employed model-agnostic techniques (e.g., LIME) for line-level defect prediction (e.g., predicting which lines will be defective in the future), helping developers to localize defective lines in a cost-effective manner.
In another example, Jiarpakdee~\ea ~\cite{jiarpakdee2020empirical} and Khanan~\ea ~\cite{khanan2020jitbot} employed model-agnostic techniques (e.g., LIME) for explaining defect prediction models, helping  developers better understand why a file is predicted as defective.
Rajapaksha~\ea ~\cite{rajapaksha2021sqaplanner}~and Pornprasit~\ea ~\cite{pornprasit2021jitline} proposed local rule-based model-agnostic techniques to generate actionable guidance to help managers chart the most effective quality improvement plans.

\subsection{Research Gaps}

While there exist research efforts on the explainability of classification tasks in SE domains (e.g., defect prediction), little research is focused on transformer-based pre-trained code models.
Particularly, practitioners often raised concerns e.g., \emph{why this source code is generated?} \emph{why this code token is modified?}.
A lack of explainability of code models could lead to a lack of trust, hindering the adoption in practice.
% To address this challenge, prior studies in the software engineering domain have employed various Explainable AI approaches on transformer-based code models ~\cite{kenny2021explaining,mohankumar2020towards,kobayashi2020attention,liu2021exploring}.
% For example, 
% Cito~\ea ~\cite{cito2021explaining} proposed an approach to generate a global explanation to understand the weaknesses of the code models.
% Cito~\ea ~\cite{cito2021counterfactual} proposed an approach to generate counterfactual explanations to explain the model's behavior by letting the end user know if the source code had been changed in a specific way, how the model’s prediction would be. This will help the users to have a more specific answer, when asking the model to do a task like security vulnerability detection.
% Also, another research using probing tasks has surprisingly claimed that CodeBERT and GraphCodeBERT which are trained on codes have a very slim difference in code understanding, compared to an NL model such as BERT ~\cite{karmakar2021pre}. Another work has replicated an NLP study ~\cite{clark2019does} on BERT, but using codes as their training benchmark. They focused on the self-attention mechanism of BERT and compared the attention behavior of the model in NLP and code.
To address this challenge, this paper aims to %analyze the pre-trained code models through the lens of Explainable AI.
%Specifically, we aim to 
address the following research questions: \emph{(RQ1) \rqtwo} and \emph{(RQ2) \rqthree}

% \begin{enumerate}[{\bf(RQ1)}]
%     \item 
%     \item 
% \end{enumerate}

% no efforts have been made to study the different ways that each code model learns to perceive the source codes when doing different downstream tasks. Do they learn to focus on different parts of the code based on their given and fine-tuned tasks? Are there meaningful differences between similar code models understanding of the code, doing the exact same task?

% \smallsection{Novelty \& Contributions.} To the best of our knowledge, this paper is the first to discover hidden patterns in different code models while they're performing certain downstream tasks and how their way of learning and understanding the codes differ, according to the tasks they are fine-tuned to do. Also, comparisons between similar code models have been made to better understand their strengths and weaknesses objectively or compared to each other. 

\section{Experimental Setup} \label{research_method}

%In this section, we discuss the design of our experiment to answer the research questions.

\subsection{Selecting Pre-Trained Code Models}

Pre-trained code models, like transformers, are deep learning models trained on extensive datasets (e.g., GitHub projects, StackOverflow posts) for source code understanding and generation. These models, also known as language models of code, employ self-supervision techniques, including BERT-based architectures like Masked Language Modeling (MLM) and Next Sentence Prediction (NSP). This training on large corpora enables them to grasp universal representations of both source code and programming-specific natural language. These models offer valuable benefits for diverse downstream tasks, eliminating the need for building new models from scratch and enhancing reusability. Notably, recent advancements have produced various Transformer-based pre-trained code models (e.g., CodeBERT, GraphCodeBERT, CodeGPT, CodeT5). This paper concentrates on two specific transformer models: CodeBERT and GraphCodeBERT.
%\kla{Hadi, can you provide justification why CodeBERT and GraphCodeBERT? //Hadi: OK I added some per item in the below} 

\textbf{CodeBERT}~\cite{feng2020codebert} is a bimodal multi-lingual pre-trained model for a programming language (PL) and natural language (NL). It has a multi-layer Transformer encoder, trained on Masked Language Modeling (MLM), and Replaced Token Detection (RTD) with both NL and PL as inputs. The model has a similar architecture as BERT ~\cite{devlin2018bert} and showed promising results on multiple downstream tasks such as Code Translation, Clone Detection, Defect Detection, etc ~\cite{feng2020codebert}. CodeBERT has been chosen as one of our code models since its popularity at the time of conducting this study and the many related works that either propose a technique based on it or use it as a comparison baseline~\cite{pan2021empirical}.

\textbf{GraphCodeBERT}~\cite{guo2020graphcodebert} is a pre-trained model similar to CodeBERT that considers the semantic-level structure of the code as well. It uses data-flow in the pre-training stage and uses MLM, alongside Edge Prediction and Node Alignment as pre-training tasks. Having this feature included, the model is able to improve the results on its benchmark tasks compared to CodeBERT, but as we will show in our experiments, in tasks like Code Document Generation, it shows a great drawback.
GraphCodeBERT has been selected as one of our code models since at least in theory, it is a step up compared to CodeBERT with information added to model from the code itself. In other words, it belongs to another category of code models, which helps with generalizability of our findings.

\subsection{Fine-tuning Pre-Trained Code Models on Three Downstream Tasks}

The existing pre-trained code models have been used for various downstream software engineering tasks. which can be categorized into four types: (1) Text$\rightarrow$Text (e.g., language translation of code documentation ~\cite{lu2021codexglue}, query reformulation ~\cite{cao2021automated}); 
(2) Text$\rightarrow$Code (e.g., code search ~\cite{gu2018deep, nguyen2016t2api});
(3) Code$\rightarrow$Text (e.g., code summarization ~\cite{haque2020improved}, commit message generation ~\cite{jiang2017automatically, liu2018neural});
and (4) Code$\rightarrow$Code (e.g., automated program repair ~\cite{jiang2021cure, li2020dlfix, chen2019sequencer}, programming language translation ~\cite{roziere2020unsupervised}, code completion ~\cite{Svyatkovskiy2020IntelliCode}). In this paper, we will focus on the following three downstream tasks.

\textbf{Code Document Generation (CDG)} or Code Summarization (Code$\rightarrow$Text) is an NLP task designed to generate natural language comments for a given source code, which could help developers to better understand codes in software projects with wrong or missing comments and decrease the extra time that should be spent on reading the source code. 
For example, given a Python method (``\texttt{def sum(x,y): ...}''), the NLP model will generate natural language comments as (``This is a summation function'').

% Currently, Transformer-based models proved to be state-of-the-art due to their capability at considering long dependencies for longer texts and source codes ~\cite{li2022setransformer}.

\textbf{Code Refinement} (Code$\rightarrow$Code) is an NLP task designed to generate refined source code (e.g., a fixed version) for a given source code (e.g., a buggy version).
Code refinement has been widely studied in the context of code review ~\cite{tufano2022using, thongtanunam2022autotransform,liu2022autoupdate}, helping developers receive refined code that is likely to be approved without waiting for reviewers' feedback.

\textbf{Code Translation} (Code$\rightarrow$Code) is an NLP task designed to generate source code in one language (e.g., Java) for a given source code in another language (e.g., C\#).

\subsection{Hyper parameter settings} 

During the model training, we use default parameter values as follows: max source length of 256 and max target length of 128 with the learning rate of 5e-4, with 16 as batch size and training it for 100 epochs.

Both models follow the same steps to generate the output. After training a model on a downstream task on the respective training dataset, the model generates the output for each test data item, token by token. That is, in the inference time, in each step, some tokens (depending on the beam size, set in the model) are chosen from the potential predicted candidates, and this process is repeated (new tokens are added to the candidate output string) until the model generates the end-of-sentence token, which indicates the end of prediction.

\subsection{Evaluating the Model Accuracy} 

To ensure that the explanations generated from our models are reliable, the models should be accurate.
To evaluate the model accuracy, we use smoothed BLEU-4 score that is used in CodeBERT's original study ~\cite{feng2020codebert} and is commonly used by baseline document generation techniques ~\cite{hu2018deep}. For other tasks, we use a BLEU-4 score.
The BLEU-4 is the only evaluation metric we use in this paper and from now on, unless we explicitly say otherwise when we use the BLEU score we are referring to the BLEU-4 score.
BLEU score~\cite{papineni2002BLEU} calculates the n-gram overlap of the prediction and the gold document or code snippet. 
% for all n-gram orders up to four. In other words, it counts the number of all n-gram (from one to four) sub-sequences of the predicted output that are also in the gold string and it gives all n-gram scores an equal weight to calculate the final score. 
Since in CDG, the generated sentences are usually short, and the higher-level n-gram is not likely to have an overlap, CodeBERT uses a smoothed version ~\cite{lin2004orange} that compensates it, by giving additional counts to higher-level n-gram overlaps.

For the code document generation task, CodeBERT achieves an overall BLEU score of 17.83 (19.06 and 17.65 per Python and Java, respectively), while GraphCodeBERT achieves an overall BLEU score of 5.3 and 4.21 for Java and Python, respectively. 
For the code refinement task, CodeBERT achieves a BLEU score of 91.07 with an exact match of 5.16, while GraphCodeBERT achieves a BLEU score of 91.31 with an exact match of 9.1.
For the code translation task (Java to C\#), CodeBERT achieves a BLEU score of 79.92 with an exact match of 59.0, while GraphCodeBERT achieves a BLEU score of 80.58 with an exact match of 59.4.

\begin{table}[t]
\caption{Dataset Statistics.}
\centering
\resizebox{\linewidth}{!}{
% \begin{center}
\renewcommand*{\arraystretch}{1.3}
\begin{tabular}{cccccc}
\hline
Task &
  \begin{tabular}[c]{@{}c@{}}Input$\rightarrow$Output\end{tabular} &
  Training &
  Validation &
  Test &
  Total \\ \hline
Code Translation & Java$\rightarrow$C\#             & \multicolumn{1}{l}{10,300} & 500        & 1,000  & 11,800 \\ \hline
\multirow{2}{*}{\begin{tabular}[c]{@{}c@{}}Code Document\\ Generation\end{tabular}} & Java$\rightarrow$NL & 164,923 & 5,183 & 10,955 & 181,061 \\
                 & Python$\rightarrow$NL             & 251,820  & 13,914     & 14,918 & 280,652 \\ \hline
Code Refinement                              & Java$\rightarrow$Java & \multicolumn{1}{l}{52,364} & 6,545  & \multicolumn{1}{l}{6,545} & 65,454 \\ \hline
\end{tabular}
% \end{center}
}
\label{dataset}

\vspace{-10pt}
\end{table}

\subsection{Explaining the Models via Attention Scores}

Transformer models are able to comprehend long dependencies among words in a sentence (or tokens in a code snippet) by benefiting from the attention mechanism. 
The attention mechanism basically works with a key ($k$), query ($q$), and value ($v$), and $d_k$ denotes the dimensionality of the key. In its simplest form, it calculates the similarity between the query, key, and values as:

$$
\operatorname{Attention}(q, k, v)=\operatorname{softmax}\left(\frac{q k^{T}}{\sqrt{d_{k}}}\right) V
$$

where keys and queries are the elements in the sequence. Calculating all the dot products of $q_i$.$k_j$ will result in a matrix where each row represents the attention weight for a specific element $i$ to all other elements in the sequence. Afterward, a softmax layer and multiplication with the value vector will be applied to the matrix to obtain the weighted mean. This means every dependency between every two elements will be considered in the final output.

Attention is a reasonable XAI method to explain CodeBERT and GraphCodeBERT. To calculate the attention per token, we need the weights for the encoder-decoder attention layers. Since we have six Transformer decoder layers stacked up, we have six attention layers. Rather than somehow aggregating these 6 layers' attention values into one metric, we decided to keep all layers' data and analyze the different layers' roles in explaining the outputs. The attention weights are available inside the model, but the Transformers library that is used by CodeBERT and GraphCodeBERT doesn't provide them by default. Hence, we changed the model's implementation to collect them, as well. Note that the attention score for each token is the output of a softmax layer, therefore it has a value between 0 and 1.

\section{Experimental Results}
\label{results}

%In this section, we present the experimental results with respect to our two research questions:

%\subsection*{\textbf{(RQ1) \rqtwo}}
\subsection{(RQ1) \rqtwo}

% \smallsection{Motivation}  
% \label{RQ2_motive}
% Applying a Transformer-based model on a source code, for any downstream task, accepts the source code snippets as sequences of tokens. Yet, little is known about what the code models learn from that code snippets. For example, what kinds of code tokens are often highlighted by the model in the learning process and whether such highlighted tokens are semantically important or not.
% In general, knowing what has been learnt is important from two perspectives: (1) Trust: depending on what tokens are impacting the output one can rely more or less on the recommendations. For example, if XAI reveals that a model makes its decision mainly based on parentheses and indentations in the source code, the user will not (and should not) trust its recommendations, (2) Debugging: knowing which tokens have the highest influence on a generated wrong output, one can devise mechanisms to change the training data, models, or the training process to fix the issue.

% Therefore, in this RQ, we study the attention weights of different token types provided by CodeBERT and GraphCodeBERT's internal layers, to see which kind of source code tokens have the highest influence on each generated output, for a given downstream task. This helps us understand whether these models' success and failure are related to focusing on the right/wrong tokens or whether the highlighted tokens are reasonable and the potential problems' root causes are somewhere else (e.g., inappropriate evaluation metrics).

\smallsection{Approach}
To answer this RQ, we % As discussed, in \ref{RQ2_motive}, to explain the models, we
analyze the attention weights per token type and not individual tokens. To do so, we first need to define a list of token types. This is a subjective decision on what tokens types are of interest for our study. We opt for a set of seven token types that cover all tokens and group them according to their semantic relevance, as follows:

\textbf{Method name:} The method under study's name can be one of the main decision factors on perceiving what a method does which is a very important step for the model, specially in some tasks such as CDG. This only includes the main method's name in each sample (reminder that each input sample is the source code for one method) and not the methods called within the main method's body.

\textbf{Type identifiers:} This category represents all the keywords that are used for identifying token types in our source languages Python and Java. For more information on these tokens, look at ``type\_identifier'' and ``*\_type'' node types in tree-sitter for each language.

\textbf{Language keywords:} Control flow command tokens are all bundled together for each language in this category. These tokens are as follows: For Python: \{\textit{False, None, True, and, as, assert, async, await, break, class, continue, def, del, elif, else, except, finally, for, from, global, if, import, in, is, lambda, nonlocal, not, or, pass, raise, return, try, while, with, yield}\} and for Java: \{\textit{if, else, switch, case, while, class, enum, interface, annotation, public, protected, private, static, abstract, final, native, synchronized, transient, volatile, strictfp, assert, return, throw, try, catch, finally, default, super, do, for, break, continue, super, void, import, extends, implements, import, instanceof, new, null, package, this, throws}\}.   
%can be found in Table~\ref{tab:syntax_tokens}. 

%\begin{table}[]
%\caption{Control flow command tokens for Java and Python.}
%\label{tab:syntax_tokens}
%\centering
%\begin{tabular}{cc}
%\hline\\
%\vspace{0.1cm}Language &
%  Tokens \\ \hline
%Python &
%  \begin{tabular}[c]{@{}c@{}}\\False, None, True, and, as, assert, async, await, break,\\ class, continue, def, del, elif, else, except, finally, for,\\ from, global, if, import, in, is, lambda, nonlocal, not,\\ or, pass, raise, return, try, while, with, yield\\\\\end{tabular} \\ \hline
%Jave &
%  \begin{tabular}[c]{@{}c@{}}\\if, else, switch, case, while, class, enum, interface, \\ annotation, public, protected, private, static, abstract, \\ final, native, synchronized, transient, volatile, strictfp, \\ assert, return, throw, try, catch, finally, default, super,\\ do, for, break, continue, super, void, import, extends, \\ implements, import, instanceof, new, null, package, \\ this, throws\\\\\end{tabular} \\ \hline
%\end{tabular}
%\end{table}

\textbf{Method calls:} This category includes all the tokens that are the name of methods invoked within the body of the method under study. They can also be instrumental in describing what the method is doing and may contain bugs to fix.

\textbf{Local variables:} 
Here we consider all variables that are used only in the body of the method under study. That is, the input arguments of the method are excluded.

\textbf{Input variables:} This category only contains the input arguments of the method under study. We have separated it from the Local variables category.

\textbf{Others:} This category represents all the tokens that are not included in any of the categories above. Mostly tokens like punctuation, constant values, parentheses, etc.

Note that although these token types are chosen subjectively, the results will justify this design choice by showing that they are among the most important tokens and not many contributing tokens are left for the ``Others'' category. We should also emphasize that the level of abstraction on what constitutes a ``token category'' is up to the XAI user. For instance, we decided to separate the Input Arguments category from the Variable Names category, to better analyze their effects individually, but merging the two categories is a valid design choice as well (just in different level of abstraction). 

For each sample in test data, we follow these steps to find the distribution of attention scores over different categories:

For each generated token (each step), we take its attention weights toward the input tokens and find their corresponding type. Then, we accumulate the attention scores of all the tokens in each category to get a total score for that category in that sample. Our categories cover all tokens so the sum of all scores for each step is equal to one.
We go through the same process for all output tokens in all code snippets of the testing dataset and gather the accumulation of attention scores for each category.

It's noteworthy that the size of token categories is quite imbalanced. For example, there is only one method name for each sample but many tokens in the ``others'' category. Therefore, we normalize the total score of each category, according to its population as in Table~\ref{tab:population_of_categories}. This gives us the attention score per token for each category. Finally, we normalized scores of all categories, between 0 and 100 for the purpose of easier comparison.

\begin{table}[]
\centering
\caption{Number of tokens in each category (1: CT, 2: CDG\_Java, 3: CDG\_Python, 4: CR) for each task}
\label{tab:population_of_categories}
\renewcommand*{\arraystretch}{1.4}
\scalebox{0.92}{
\setlength{\tabcolsep}{2.5pt}
\begin{tabular}{ccccccccc}
\hline
\multicolumn{1}{l}{} &
  \renewcommand*{\arraystretch}{1.1}\begin{tabular}[c]{@{}c@{}}Method \\ name\end{tabular} &
  \renewcommand*{\arraystretch}{1.1}\begin{tabular}[c]{@{}c@{}}Input\\ variables\end{tabular} &
  \renewcommand*{\arraystretch}{1.1}\begin{tabular}[c]{@{}c@{}}Method\\ call\end{tabular} &
  Variable &
  \renewcommand*{\arraystretch}{1.1}\begin{tabular}[c]{@{}c@{}}Type\\ identifier\end{tabular} &
  \renewcommand*{\arraystretch}{1.1}\begin{tabular}[c]{@{}c@{}}Language\\ keywords\end{tabular} &
  Others &
  Total \\ \hline
1         & 1,033  & 2,859   & 2,065  & 4,289   & 2,815  & 3,408  & 21,731  & 38,200  \\ \hline
2   & 12,991 & 44,175  & 42,285 & 101,352 & 63,143 & 73,644 & 446,419 & 784,009 \\ \hline
3 & 16,633 & 110,630 & 19,567 & 197,878 & 0      & 77,662 & 487,860 & 910,230 \\ \hline
4         & 7,761  & 19,316  & 33,717 & 70,750  & 52,969 & 36,408 & 319,544 & 540,465 \\ \hline
\end{tabular}
}
\vspace{-10pt}
\end{table}

Among these defined categories, ``Method name'', ``Input variables'', and ``Local variables'' are more representative of the naming aspects of a source code. So we group them in a higher-level category of ``Naming'', while ``Method calls'', ``Type identifiers'', and ``Language keywords'' are more related to the structure of the code. Thus, we consider them as the higher-level category of ``Structure''. Also note that we only report the average score of all six layers for each task and model here, since reporting all results per layer would be too lengthy and also the results of this RQ were quite similar over different layers and they all followed the same patterns.

\smallsection{Results}
\label{RQ2_results}
In the three downstream tasks under study, we expect different normalized attention scores per high-level category, as follows: (a) CT is a task that heavily relies on the structure, since the model must learn the source language structure, and generate the equivalent structure in the target language. (b) In CDG, the structure is less important (nested blocks and syntax trees have less to do with the output document). On the other hand, names are very important in this task, since they basically describe the functionality of the source code. (c) Finally, we expect code refinement to be in the middle of these two ends, since both names and the structure are important in debugging a code.

\begin{table}[]
\centering
\caption{Normalized attention score of the two high-level categories of tokens, for different code models and tasks. The results are the average of all six layers for each task.}
\label{tab:RQ2_high_level}
\renewcommand*{\arraystretch}{1.4}
\begin{tabular}{clccc}
\hline
Task & \multicolumn{1}{c}{Model} & \multicolumn{1}{c}{Naming} & \multicolumn{1}{c}{Structural} & \multicolumn{1}{c}{Others} \\ \hline
\multirow{2}{*}{Code Translation} & CodeXGLUE     & 42.36\% & 51.38\% & 6.27\% \\ \cline{2-5} 
                                  & GraphCodeBERT & 42.60\% & 51.30\% & 6.09\% \\ \hline
\multirow{2}{*}{CDG\_java}        & CodeXGLUE     & 63.31\% & 29.19\% & 7.50\% \\ \cline{2-5} 
                                  & GraphCodeBERT & 64.51\% & 28.19\% & 7.30\% \\ \hline
\multirow{2}{*}{CDG\_python}      & CodeXGLUE     & 67.97\% & 23.75\% & 8.28\% \\ \cline{2-5} 
                                  & GraphCodeBERT & 74.63\% & 17.83\% & 7.54\% \\ \hline
\multirow{2}{*}{Code Refinement}  & CodeXGLUE     & 56.46\% & 37.96\% & 5.58\% \\ \cline{2-5} 
                                  & GraphCodeBERT & 55.49\% & 38.86\% & 5.65\% \\ \hline
\end{tabular}
\vspace{-10pt}
\end{table}

Table~\ref{tab:RQ2_high_level} shows the normalized total attention score of each high-level category and validates our hypothesis. Code Document Generation, as a task that heavily relies on naming, has a considerably high normalized attention score of over 63\% for the Naming category, while it pays much less attention to the Structural token types category, compared to other tasks. It has also a higher number for the Other category which can be understandable considering the fact that NL comments in the code are also part of this category. 
Code Translation, on the other hand, is the only task that has more than 50\% normalized attention score for Structural tokens and less than any task for the Naming category. Code Refinement in this comparison holds the middle ground between the two mentioned tasks in both categories.

In Table~\ref{tab:RQ2_details}, we have a more detailed analysis for each of our categories. We saw that both models pay more attention to the structural tokens for CT. Getting into more details, the results show that this attention is more focused on Method calls, and Type identifiers rather than Language keywords, roughly having only 6\% of the normalized score. It is very interesting that studying individually, the Method name category has the second highest score after Method calls. Even though in translating a code, the method name is usually unchanged, this shows that \textbf{the models considerably utilize the name of the method to understand its functionality.}

In the CDG task, we observed a great reliance on Naming categories, the results show that the Method name category plays the most significant role. It always has a normalized score of close to 40\% or higher.  In the absence of Type identifiers, in GraphCodeBERT CDG\_python, this category has the highest score ever among all the token types across all tasks/models. Intuitively, this amount of importance is justifiable, given that even humans rely a lot on the method names to understand their functionality. In addition, in three out of four different experiments for this task, the Input variables have the second highest normalized attention score (between 11\% to 16\%). This basically means that \textbf{the models have learned that while generating document for a method, the main and most interesting part is the signature of the method and not the body.}

Code Refinement which holds a middle ground between two other tasks has a very close score for four categories of Method name, Input variable,  Method call, and Variable. Since in this task the model is supposed to look for bugs and try to fix them, it seems that \textbf{the models have learned that fewer bugs happen in categories like Type identifier, Language keywords, and Others.} This makes sense because we know that databases for this task are gathered from public projects and they probably do not have syntactical errors. The method name is also less likely to have a bug, but as we saw in other tasks, this category always has a minimum appeal for the models.

It is interesting that according to the table, Type identifiers which is a category purely related to the syntax of the code and do not include any naming, have the most contribution to CT with a gap compared to others. It has a score of 20.67\% and 21.66\% for CodeBERT and GraphCodeBERT respectively while its score in other categories is below 14\%. Also, it is worth mentioning that GraphCodeBERT which uses the dataflow of the code in order to capture its structure; always has a higher score for this category compared to CodeBERT.

The same patterns appear to be valid for Variable names and for CR. The score of this category for CR is 17.85\% and 18.93\%, while in other tasks the score is always below 12\%. 

Similarly, the Method name category has a harshly higher score in CDG. For this task in python, this category has a score of 39.44\% and 41.04\%, and in Java, it has 46.17\% and 54.21\% for CodeBERT and GraphCodeBERT, respectively.

\begin{table}[]
\centering
\caption{Normalized attention score of different token types, for different code models and tasks. The results are the average of all six layers for each task.}
\label{tab:RQ2_details}
\renewcommand*{\arraystretch}{1.4}
\scalebox{0.8}{
\setlength{\tabcolsep}{2.5pt}
\begin{tabular}{cccccccc}
\hline
Task/Model                                            & \renewcommand*{\arraystretch}{1.2}\begin{tabular}[c]{@{}c@{}}Method \\ name\end{tabular} & \renewcommand*{\arraystretch}{1.2}\begin{tabular}[c]{@{}c@{}}Input\\ variables\end{tabular} & \renewcommand*{\arraystretch}{1.2}\begin{tabular}[c]{@{}c@{}}Method\\ call\end{tabular} & Variable & \renewcommand*{\arraystretch}{1.2}\begin{tabular}[c]{@{}c@{}}Type\\ identifier\end{tabular} & \renewcommand*{\arraystretch}{1.2}\begin{tabular}[c]{@{}c@{}}Language\\ keywords\end{tabular} & Others \\ \hline
\begin{tabular}[c]{@{}c@{}}CT-Code\\ XGLUE\end{tabular}     & 21.36\%                                                & 9.63\%                                                    & 24.26\%                                               & 11.36\%  & 20.67\%                                                   & 6.45\%                                                      & 6.27\% \\ \hline
\begin{tabular}[c]{@{}c@{}}CT-Graph\\ CodeBERT\end{tabular} & 22.78\%                                                & 7.89\%                                                    & 23.46\%                                               & 11.93\%  & 21.66\%                                                   & 6.18\%                                                      & 6.09\% \\ \hline
\begin{tabular}[c]{@{}c@{}}CDG\_Java-Code\\ XGLUE\end{tabular}     & 39.44\%                                                & 13.88\%                                                   & 10.49\%                                               & 10.00\%  & 13.07\%                                                   & 5.63\%                                                      & 7.50\% \\ \hline
\begin{tabular}[c]{@{}c@{}}CDG\_Java-Graph\\ CodeBERT\end{tabular} & 41.04\%                                                & 15.10\%                                                   & 8.44\%                                                & 8.38\%   & 13.13\%                                                   & 6.62\%                                                      & 7.30\% \\ \hline
\begin{tabular}[c]{@{}c@{}}CDG\_Python-Code\\ XGLUE\end{tabular}   & 46.17\%                                                & 11.96\%                                                   & 16.00\%                                               & 9.83\%   & 0.00\%                                                    & 7.76\%                                                      & 8.28\% \\ \hline
\begin{tabular}[c]{@{}c@{}}CDG\_Python-Graph\\ CodeBERT\end{tabular} & 54.21\%                                                & 12.79\%                                                   & 10.79\%                                               & 7.64\%   & 0.00\%                                                    & 7.03\%                                                      & 7.54\% \\ \hline
\begin{tabular}[c]{@{}c@{}}CR-Code\\ XGLUE\end{tabular}     & 22.01\%                                                & 16.60\%                                                   & 20.33\%                                               & 17.85\%  & 9.82\%                                                    & 7.81\%                                                      & 5.58\% \\ \hline
\begin{tabular}[c]{@{}c@{}}CR-Graph\\ CodeBERT\end{tabular} & 19.36\%                                                & 17.19\%                                                   & 21.15\%                                               & 18.93\%  & 10.02\%                                                   & 7.69\%                                                      & 5.65\% \\ \hline
\end{tabular}
}
\vspace{-10pt}
\end{table}

% \finding{\textbf{Answer to RQ2:}
Having all these observations, we can see a pattern of importance comparing different tasks together. The method name and input variable (basically the first line of the code samples) are the most important categories for CDG; Method calls and local variables play the most significant role on code refinement, alongside the method name with slightly lower importance. On the other hand, code translation is concerned with type identifiers and language keywords, more than any other task, while still caring about some naming categories, as well. In Table~\ref{tab:RQ2_high_level}, we have aggregated the numbers for our two main categories and we can see a pattern that we expected. Naming tokens are important for all tasks, but less for code translation which alternatively, cares more about structural tokens compared to other tasks.
% }

\subsection{(RQ2) \rqthree}

% \smallsection{Motivation}  
% In RQ1, we made sure attention is a reasonable tool to explain CodeBERT and GraphCodeBERT, and in RQ2 we show that the tokens are mostly picked up correctly by the model, so the models learn the right tokens per task, in most cases. In this RQ, we delve deeper and explore the scenarios where the code models did not perform well but the problem at hand was not that difficult. Answering this question will help understand what should be done to make the model at least perform consistently well, on simple data items.

\smallsection{Approach} \label{RQ3_design}
To provide explanations on when CodeBERT and GraphCodeBert perform well and when they fail, in this RQ, we start by a qualitative analysis of some sample predictions. Then we make hypotheses based on our observations and finally verify them quantitatively on the whole dataset. To define the strong and weak performances of the models, we cannot simply rely on the absolute values of the evaluation metric (BLEU). Since the magnitude of the BLEU score partially depends on how difficult or easy the document generation task is, per sample code. Therefore, we need to somehow measure the difficulty level of the document generation task, given a source code.

To address this research question, we create metrics to gauge sample complexity and evaluate model performance accordingly.

In the context of CR and CT, where token copying predominates, Levenshtein Distance (LD) between input and expected output ('gold output') serves as a suitable complexity metric. By calculating LD for all dataset samples, we identify the easiest third based on lower LDs.

For the CDG task, despite input-output disparity in programming language (PL) and natural language (NL), our approach remains similar. We determine difficulty by intersecting preprocessed tokens in the gold output document per method with tokens in the method's source code.

To find the overlap between source code and output tokens, we follow these preprocessing steps: First, we remove punctuations and tokens shorter than three characters, in the output document. Then, lemmatize those tokens using the standard Wordnet ~\cite{NLTK} lemmatizer, offered in the NLTK  package. Next, we tokenize the source code using a parser for the respective language. Note that 
% as explained in Section~\ref{preprocessing}, 
due to CodeBERT and GraphCodeBERT's tokenizations, which may split one meaningful word into multiple tokens, we do not use their tokenization for this analysis. Finally, we create a set of case-insensitive tokens that fall at the intersection of processed output and source code tokens.

Now an easy/difficult document generation task is when the overlap between the two sets is high/low.
Therefore, the same as the cut-offs for CR and CT, we consider the first one-third samples with the highest overlaps as easy, and the one-third cases with the least overlaps as hard problems, and ignore the rest (average difficulty-level).

The above process tells us which samples are considered hard and which ones are considered easy. Now we need to measure the performance of models. To do so, we use BLEU score since it is the most accepted and reliable score that is applied in these tasks in the literature. For the accuracy, we also take the one-third of the samples with the highest BLEU scores as $High$ and the samples with the one-third least BLEU scores as $Low$.

Having these definitions, there will be four categories (for the tuples of $<$easiness level, model accuracy$>$) as below:

\textbf{$Easy-High$:} 
This category contains test data items that are easy problems (meaning high similarity between the input and the gold data) that the models have achieved a High BLEU score on them. 

\textbf{$Hard-High$:} 
This category contains the samples labeled as $Difficult$ and $High$. This means even with the lack of common tokens, the model was able to achieve a satisfactory result in these cases.

\textbf{$Hard-Low$:} 
This category includes the cases that are $difficult$ again, and expectedly, end up with $Low$ accuracy for the model's prediction.

\textbf{$Easy-Low$:} 
This category is the most interesting one in this paper, since it can show the potential weaknesses of the model and is very suitable for being analyzed and ``explained''. 
Samples in this group, are among the samples with higher overlaps in their corresponding dataset that means that the model is having rather an easy job predicting. However, the BLEU score as our indicator of the model's accuracy is showing poor performance comparing to other samples.

In order to perform our manual observation (qualitative study) for this RQ, after grouping our test dataset into these four categories, we randomly choose 100 samples from our target category ($Easy-Low$), and manually analyze their outputs and attention weights. 

For each sample, we record the most interesting findings to identify the most frequent patterns. This way we develop some hypotheses. Finally, we try to verify these hypotheses by quantitatively studying the whole test dataset, with respect to the hypotheses. The output of this quantitative phase is in the form of some descriptive statistics to either confirm or reject the observations made based on the 100 samples. 

\smallsection{Results}
Having the data divided according to the defined groups, Table~\ref{tab:easy-bad-ratio} shows the ratio of the target category population to the whole dataset for each task-model.
%and also, Fig.~\ref{fig:RQ3_populations} and Fig.~\ref{fig:RQ3_populations_CDG} show the distribution of the target category. 
Next, we will explain the observations from the manual analysis.

\begin{table}[]
\centering
\caption{The ratio of the Easy-Low category population to the whole dataset, for each task-model.}
\vspace{-10pt}
\label{tab:easy-bad-ratio}
\renewcommand*{\arraystretch}{1.4}
\begin{tabular}{ccc}
\\ \hline
\multicolumn{1}{c}{} & CodeBERT & GraphCodeBERT \\ \hline
CT   & 11.70\%  & 11.41\%       \\ \hline
CDG / Java           & 5.12\%  & 5.59\%       \\ \hline
CDG / Python         & 5.87\%  & 5.85\%       \\ \hline
CR      & 2.49\%   & 4.37\%        \\ \hline
\end{tabular}
\end{table}

\textbf{Observations 1: The pre-trained code models do not work well, when the output gold document is long.}

\begin{figure}[tbp]
\centering
\centerline{\includegraphics[scale=0.5]{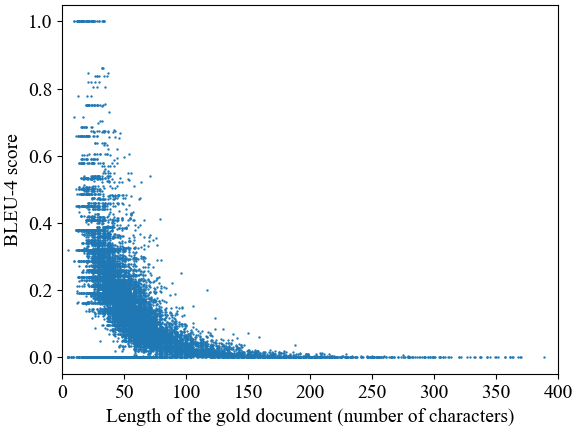}}

\caption{BLEU scores vs. the gold document code length.}
\label{BLEU-to-gold-length}
\vspace{-10pt}
\end{figure}

Our first observation regarding the CDG task is that in cases with long gold documents the BLEU scores tend to be low! We plotted the distribution of BLEU scores, according to the gold document's length, in Fig.~\ref{BLEU-to-gold-length}. According to the plot, most high BLEU scores happen when the length of the gold document is less than 50 characters.

In general, the idea of the BLEU score is about counting the number of n-grams that are common between the reference and the output. Model-generated documents are usually short so for longer sentences, there is less chance that the model chooses the same phrasing and words with the same order. Another plausible explanation for this observation is that a longer document means the method implements a more complex task and thus it is harder for the model to generate the right documentation for the complex method.

\begin{figure}[tbp]
\centering
\centerline{\includegraphics[scale=0.6]{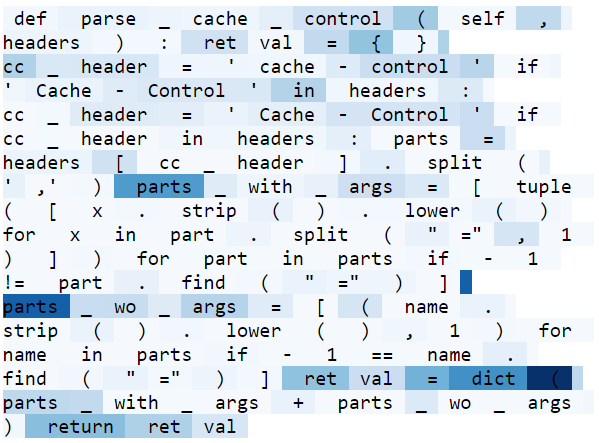}}
\captionsetup{%
    justification=justified,%
}
\begin{lstlisting}[frame=b, backgroundcolor = \color{white}]
((*\textbf{Gold document:} Parse the cache control headers returning a*))
((*dictionary with values for the different directives .*))
((*\textbf{Best prediction:} Parse a dict of headers*))
((*\textbf{BLEU score:} 0.08*))
((*\textbf{Overlap:} 0.54*))
\end{lstlisting}
\caption{A sample method broken into tokens, where the attention values of the last layer of CodeBERT for the last generated token(``headers'' in this example) are highlighted as shades of blue (the darker, the higher).}
\label{sample1_for_long_gold_document}
\end{figure}

One potential solution for this problem is forcing the model to generate longer sequences as the output which will increase the chance of a high BLEU score, in cases with long reference documents. But obviously, since this is not actually the model's fault and in these cases, a low BLEU score does not necessarily indicate a bad prediction (like the example shown in Fig.~\ref{sample1_for_long_gold_document}), the best way to handle this problem is considering other evaluation metrics and ideally more subjective ones.

\textbf{Observation 2: The pre-trained code models do not work well, when the input source code is complex.} 

\begin{figure}[tbp]
\centering
\centerline{\includegraphics[scale=0.5]{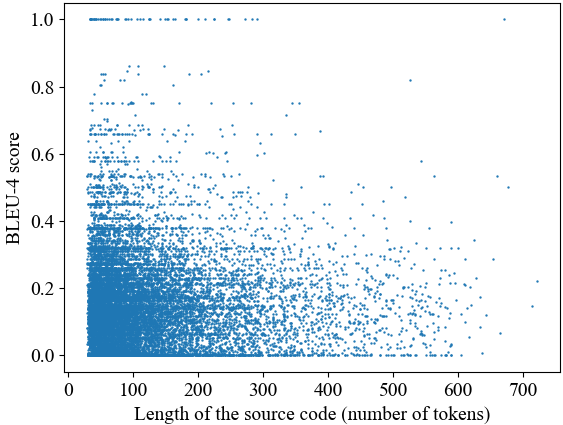}}
\caption{BLEU scores vs. the source code length.}
\label{source_length-BLEU}
\vspace{-10pt}
\end{figure}

Another interesting observation is about the length of the source code. The results show that in cases with longer code, the BLEU score is usually lower. We started the initial analysis with the CDG data and the results, which are summarized in Fig.~\ref{source_length-BLEU}, show a decreasing trend of BLEU scores, by the increase of the source code length. For example, the average BLEU score for cases shorter and longer than 300 tokens is 0.161 and 0.149, respectively.

There are two explanations for this observation: (1) the fixed length of the inputs in models, which basically means if the source code's length is higher than a fixed value (in our experiments, 256 tokens), then the input will be truncated. This means some tokens will not make it to the decoder, and thus we have a potential degradation in the final scores. (2) the increased complexity of the code. Similar to Observation 1, longer source code means more complex logic, more objects, and functionalities to consider for the model which probably leads to poorer results.

According to these reasons, two basic solutions can be suggested: (a) increasing the input threshold and (b) decreasing the input's length. The input threshold can be easily modified in the training process of the model and only requires more resources. The second option, however, is a more complex solution that is already kind of naively implemented by the truncation. Another potential alternative is to refactor long methods to multiple smaller methods, then pass each method to the models to generate documents for, and finally merge all output documents as one document. 

Next, to expand this observation on all tasks-models, and find more statistical results, we used some common code complexity metrics and conducted an analysis on all 8 model-tasks to get a better understanding of the root cause of the poor results for long code snippets. We chose 'number of tokens', 'cyclomatic complexity', 'nested block depth', 'number of variables' as complexity metrics. To measure the difficulty of the task, we also included the same 'Levenshtein distance' for CT and CR and 'overlap' for CDG, in our analysis.

For each task-model, we have five different metrics to study, so we have one plot per metric. In each plot, the distribution of samples in the respective dataset, regarding that metric is shown with blue bars, and the same distribution but only for the target category (Easy-Low) is shown in red. With this visualization, we can identify any difference in terms of trends on a specific metric in the target category vs. the whole dataset.

\begin{figure}[h]
\centering
\begin{subfigure}{.4\textwidth}
  \centering
  \includegraphics[width=\textwidth]{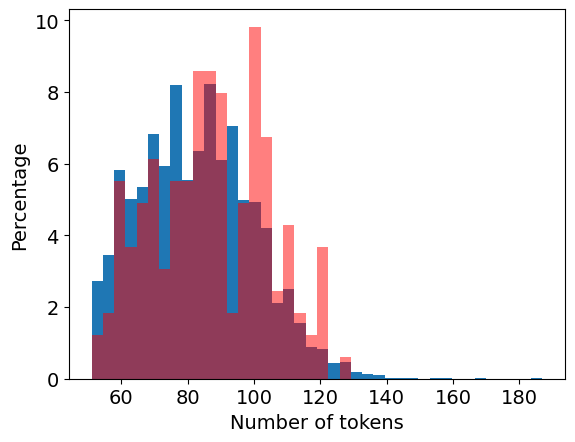}
  \caption{}%Distribution of dataset and target category according to the number of tokens}
  \label{fig:RQ3_plot_20_n_tokens}
\end{subfigure}

\begin{subfigure}{.4\textwidth}
  \centering
  \includegraphics[width=\textwidth]{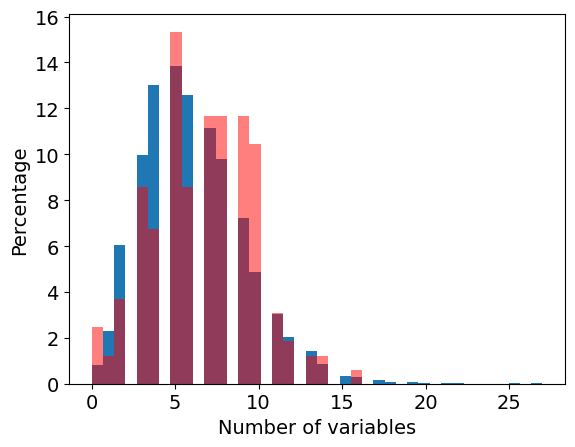}
  \caption{}%Distribution of dataset and target category according to the number of variables}
  \label{fig:RQ3_plot_23_variable}
\end{subfigure}

\caption{Distribution of the whole dataset (in blue) and the Easy-Low category (in red), according to code complexity metrics, for code refinement on CodeBERT}
\label{fig:RQ3_plot_CodeBERT_CR}
\vspace{-10pt}
\end{figure}

Fig.~\ref{fig:RQ3_plot_CodeBERT_CR} 
% and Fig.~\ref{fig:RQ3_plot_Graph_CR}
shows the results for code refinement (CR) for CodeBERT for number of tokens and the number of variables (all the plots for GraphCodeBERT and other metrics for CodeBERT can be found in the public repo). As illustrated in the plots, the Easy-Low category has a very similar distribution to the whole dataset, except for a slightly increasing trend for some reported metrics like the number of tokens and the number of variables This means that the model tends to make bad decisions, whenever the source code gets more complex in terms of the number of tokens and variables, even if the overlap of tokens is high. For instance, considering the number of tokens, as a measure of code complexity, the proportion of samples with more than 100 tokens is mostly higher in the Easy-Low category than the share of the same samples in total. It means that the samples with many tokens are more likely to be assumed ``Easy'' in our categories (more overlaps between input and outputs) but in fact, they are harder for the model to understand (given the long length of the code snippet).

% \begin{figure}
% \centering

% \begin{subfigure}{.4\textwidth}
%   \centering
%   \includegraphics[width=\linewidth]{RQ3_plot_8_n_tokens.png}
%   \caption{}%Distribution of dataset and target category according to the number of tokens}
%   \label{fig:RQ3_plot_8_n_tokens}
% \end{subfigure}

% \begin{subfigure}{.4\textwidth}
%   \centering
%   \includegraphics[width=\linewidth]{RQ3_plot_11_variable.png}
%   \caption{}%Distribution of dataset and target category according to the number of variables}
%   \label{fig:RQ3_plot_11_variable}
% \end{subfigure}
% % \hspace{-5mm}
% \caption{Distribution of the the whole dataset (in blue) and the Easy-Low category (in red), according to code complexity metrics, for code refinement on GraphCodeBERT}
% \label{fig:RQ3_plot_Graph_CR}
% \end{figure}

\begin{figure}
\vspace{-5pt}
\centering
\begin{subfigure}{.4\textwidth}
  \centering
  \includegraphics[width=.90\linewidth]{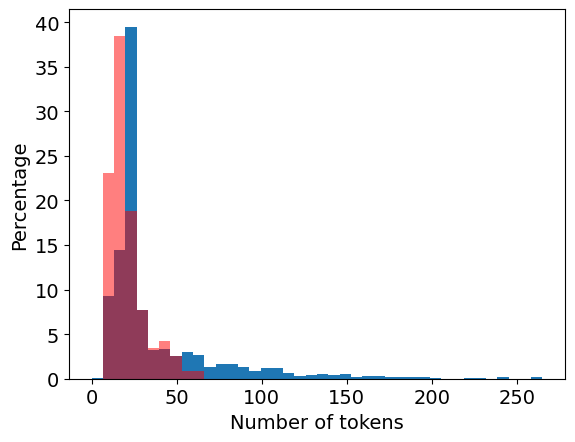}
  \caption{}%Distribution of dataset and target category according to the number of tokens}
  \label{fig:RQ3_plot_14_n_tokens}
\end{subfigure}
% \hspace{-5mm}
\begin{subfigure}{.4\textwidth}
  \centering
  \includegraphics[width=.90\linewidth]{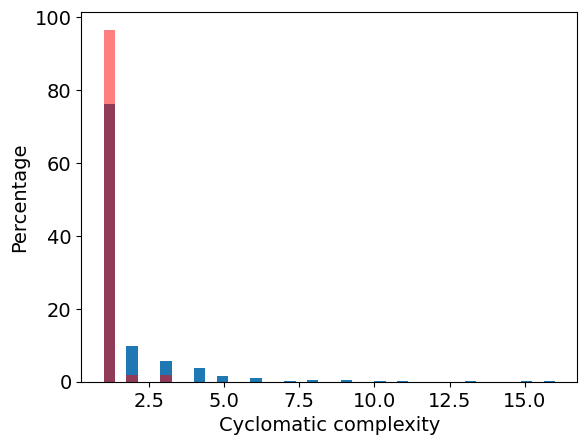}
  \caption{}%Distribution of dataset and target category according to the number of tokens}
  \label{fig:RQ3_plot_15_cyclomatic_complexity}
\end{subfigure}
% \hspace{-5mm}
\begin{subfigure}{.4\textwidth}
  \centering
  \includegraphics[width=.90\linewidth]{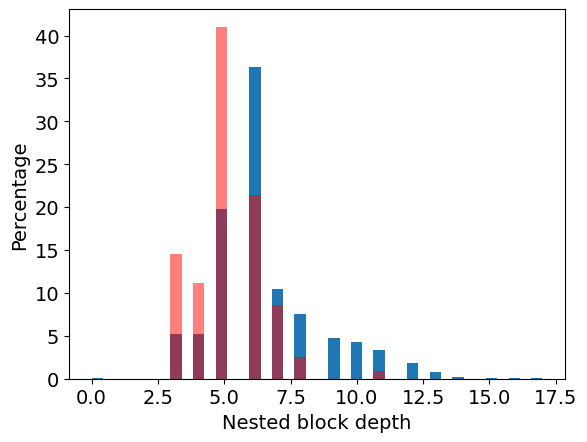}
  \caption{}%Distribution of dataset and target category according to the nested block depth}
  \label{fig:RQ3_plot_16_nested_block_depth}
\end{subfigure}
% \hspace{-5mm}
\caption{Distribution of the whole dataset (in blue) and the Easy-Low category (in red), according to code complexity metrics, for code translation on CodeBERT}
\label{fig:RQ3_plot_CodeBERT_CT}
\vspace{-10pt}
\end{figure}

% Fig.~\ref{fig:RQ3_plot_Graph_CT} and 
Fig.~\ref{fig:RQ3_plot_CodeBERT_CT} 
shows the results for code translation (CT) for the number of tokens, nested block depth, and cyclomatic complexity for CodeBERT (all the plots for GraphCodeBERT and other metrics for CodeBERT can be found in the public repo). For this task, 
number of variables follow the same pattern as the CR task, that is Easy-Low category is harder based on those metrics. On the other hand, considering the number of tokens, nested block depth, and even cyclomatic complexity, there is a reversed connection. In other words, samples that have smaller values of these metrics, have higher density in Easy-Low category. For instance, in both models, samples with tokens less than 20, are around 50\% of the target category population, even though they occupy a very small portion of the total dataset. 
One plausible explanation is that when the source code is too short (very small number of tokens and very few nested blocks), the model fails to translate it properly, due to the lack of enough information/context. Another explanation is the fact that it's harder to maintain a high BLEU score when the input is very short.

\begin{figure}

\centering

\begin{subfigure}{.4\textwidth}
  \centering
  \includegraphics[width=.90\linewidth]{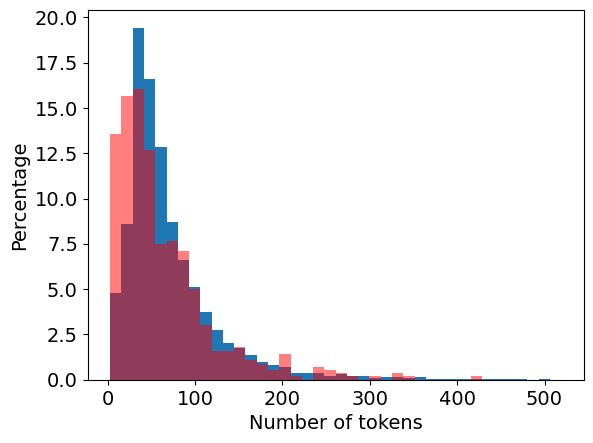}
  \caption{}%istribution of dataset and target category according to the number of tokens}
  \label{fig:RQ3_plot_26_n_tokens}
\end{subfigure}

% \begin{subfigure}{.35\textwidth}
%   \centering
%   \includegraphics[width=\linewidth]{RQ3_plot_27_cyclomatic_complexity.png}
%   \caption{}%istribution of dataset and target category according to the number of tokens}
%   \label{fig:RQ3_plot_27_cyclomatic_complexity}
% \end{subfigure}

% \begin{subfigure}{.35\textwidth}
%   \centering
%   \includegraphics[width=\linewidth]{RQ3_plot_28_nested_block_depth.png}
%   \caption{}%istribution of dataset and target category according to the nested block depth}
%   \label{fig:RQ3_plot_28_nested_block_depth}
% \end{subfigure}

\begin{subfigure}{.4\textwidth}
  \centering
  \includegraphics[width=.90\linewidth]{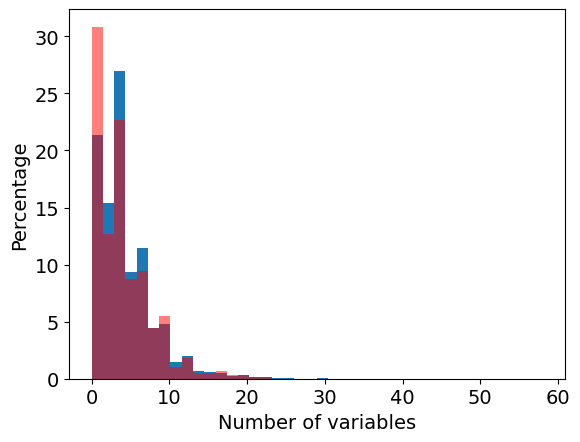}
  \caption{}%istribution of dataset and target category according to the number of variables}
  \label{fig:RQ3_plot_29_variable}
\end{subfigure}
% \hspace{-5mm}
\caption{Distribution of the whole dataset (blue) and the Easy-Low category (red), based on code complexity metrics, for code document generation on CodeBERT on Java}
\label{fig:RQ3_plot_CodeBERT_CDG_Java}
\vspace{-10pt}
\end{figure}

\begin{figure}
\centering
\vspace{-10pt}
\begin{subfigure}{.40\textwidth}
  \centering
  \includegraphics[width=.90\linewidth]{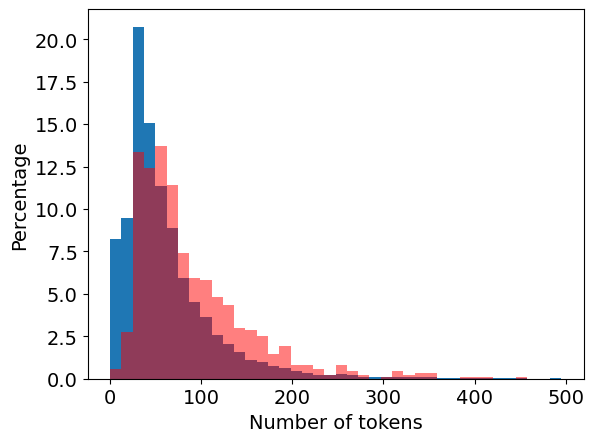}
  \caption{}%istribution of dataset and target category according to the number of tokens}
  \label{fig:RQ3_plot_32_n_tokens}
\end{subfigure}

\begin{subfigure}{.40\textwidth}
  \centering
  \includegraphics[width=.90\linewidth]{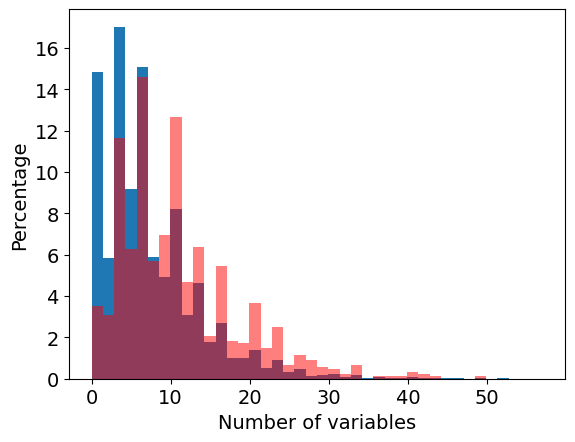}
  \caption{}%istribution of dataset and target category according to the number of variables}
  \label{fig:RQ3_plot_35_variable}
\end{subfigure}

\caption{Distribution of the whole dataset (blue) and the Easy-Low category (red), based on code complexity metrics, for code document generation on CodeBERT on Python}
\label{fig:RQ3_plot_CodeBERT_CDG_Python}
%\vspace{-5pt}
\end{figure}

Fig.~\ref{fig:RQ3_plot_CodeBERT_CDG_Java} and Fig.~\ref{fig:RQ3_plot_CodeBERT_CDG_Python} illustrate some of the similar results for code document generation (CDG) for CodeBERT (all the plots for GraphCodeBERT and other metrics for CodeBERT can be found in the public repo). In this downstream task, we can see patterns more dependent on the language, rather than the model. In all cases, the number of variables, number of tokens, and nested block depth follow the same general pattern.

In these experiments, we see that both models struggle with samples with low complexity in Java. However, they have problems figuring out the more complex samples in Python, too. For instance in Python, samples with more than 80 tokens or 8 variables, have always a higher density in the Easy-Low category compared to all of the dataset.
Finally, it's interesting that considering the cyclomatic complexity, both models in both languages struggle with samples with higher complexity.

So all-in-all, one can conclude that code models perform poorly on the code snippets with extreme values of complexity-related metrics on either direction (i.e., both long code with many nested blocks and tokens and also very short code with only a few variables and tokens).

\textbf{Observation 3: The pre-trained code models do not work well, when the models fail to focus on important categories}: 

Finally, we analyzed the contribution of token categories similar to RQ1, but specifically for the target category of ($Easy-Low$). In Table~\ref{tab:RQ3_for_target_samples} we have the normalized score of two main categories for the target samples. We were interested to compare the results for this category and the previous results for the whole test dataset. Hence, we calculated the difference between these two from Table~\ref{tab:RQ3_for_target_samples} and Table\ref{tab:RQ2_high_level}, and the outcomes are shown in Table~\ref{tab:RQ3_difference}. The negative numbers in this table indicate a decrease in the target category.

As the results show, there is a considerable decrease in the scores for the Structural category in CT. While answering to RQ1, we showed that this task mostly relied on this group of tokens  \ref{RQ2_results}. Likewise, we have a slighter decrease in the score of the Naming category in CDG while the naming category also proved to be the more important category for CDG.

Both of these clues, lead us to the conclusion that the results are less satisfying, whenever the model fails to pay enough attention to the corresponding important token category for a specific task. Based on this observation, potential research questions to investigate in the future are: ``Will the model work better if we help it by tagging the token types? Can manually amplifying the attention scores of specific categories according to the task beneficial for the code models''.

\begin{table}[]
\centering
\caption{Normalized attention score of the Easy-Low samples, in three general categories of tokens, for different code models and tasks. The results are the average of all six layers for each task.}
\label{tab:RQ3_for_target_samples}
\renewcommand*{\arraystretch}{1.4}
\begin{tabular}{ccccc}
\\ \hline
Task & \multicolumn{1}{c}{Model} & \multicolumn{1}{c}{Naming} & \multicolumn{1}{c}{Structural} & \multicolumn{1}{c}{Others} \\ \hline
\multirow{2}{*}{CT} & CodeXGLUE     & 48.85\% & 44.51\% & 7.70\% \\ \cline{2-5} 
                                  & GraphCodeBERT & 49.51\% & 43.49\% & 7.57\% \\ \hline
\multirow{2}{*}{CDG\_java}        & CodeXGLUE     & 60.08\% & 33.28\% & 8.89\% \\ \cline{2-5} 
                                  & GraphCodeBERT & 63.85\% & 37.06\% & 7.43\% \\ \hline
\multirow{2}{*}{CDG\_python}      & CodeXGLUE     & 66.37\% & 29.14\% & 8.61\% \\ \cline{2-5} 
                                  & GraphCodeBERT & 73.68\% & 24.16\% & 7.87\% \\ \hline
\multirow{2}{*}{CR}  & CodeXGLUE     & 56.64\% & 46.29\% & 5.63\% \\ \cline{2-5} 
                                  & GraphCodeBERT & 55.82\% & 48.35\% & 5.62\% \\ \hline
\end{tabular}
\vspace{-5pt}
\end{table}

\begin{table}[]
\vspace{-10pt}
\centering
\caption{The difference of normalized attention scores for the Easy-Low samples and all samples, in two general categories of tokens, for different code models and tasks. The results are the average of all six layers for each task.}
\label{tab:RQ3_difference}
\renewcommand*{\arraystretch}{1.4}
\begin{tabular}{ccccc}
\\ \hline
Task                              & \multicolumn{1}{c}{Model} & Naming  & Structural & Others  \\ \hline
\multirow{2}{*}{CT} & CodeXGLUE                 & 6.49\%  & -7.92\%    & 1.43\%  \\ \cline{2-5} 
                                  & GraphCodeBERT             & 6.90\%  & -8.38\%    & 1.48\%  \\ \hline
\multirow{2}{*}{CDG\_java}        & CodeXGLUE                 & -3.24\% & 1.85\%     & 1.39\%  \\ \cline{2-5} 
                                  & GraphCodeBERT             & -0.66\% & 0.53\%     & 0.13\%  \\ \hline
\multirow{2}{*}{CDG\_python}      & CodeXGLUE                 & -1.59\% & 1.27\%     & 0.33\%  \\ \cline{2-5} 
                                  & GraphCodeBERT             & -0.96\% & 0.63\%     & 0.33\%  \\ \hline
\multirow{2}{*}{CR}  & CodeXGLUE                 & 0.18\%  & -0.23\%    & 0.06\%  \\ \cline{2-5} 
                                  & GraphCodeBERT             & 0.33\%  & -0.30\%    & -0.03\% \\ \hline
\end{tabular}
\vspace{-10pt}
\end{table}

\section{Limitations}
\label{limitations}
One limitation of this study is that our experiments are only on Java and Python code. Including other datasets (as well as Python code for CT and CR tasks) requires lots of pre-processing to become consistent with our designs and requirements and would go beyond one conference paper's scope and size. We plan to extend these analyses to other languages in the future.

Another limitation is we used the BLEU score as our evaluation metric for the accuracy of the model, which is commonly used in document generation downstream tasks to reduce the subjectivity of the results. However, as we mentioned in the paper, it is not a comprehensive metric as it is unable to find rephrasings or cases that the prediction is not wrong, but doesn't exactly match with the gold label. 

The observations we made are also limited to the main patterns we have observed in the 100 samples, manually. Although we later quantitatively validate them, it is of course possible that there exist some other explanations as well that we have missed observing, due to the samples we have chosen. 

%We have also subjectively selected a set of six token types that we believed could have the highest impact on our models. Although, the RQ2 results showed that these categories actually contributed the most out of all tokens; the level of abstraction (grouping of tokens into categories) is a matter of design choice, which may reveal other findings (at different levels). In other words, although our observations in this level of abstraction are correct, they are not the only way to look at the tokens' attention scores and other categorizations can be studied in the future. 

In addition, the study is only limited to three downstream tasks (CT, CDG, and CR) and two code models (CodeBERT and GraphCodeBERT). More work is required to generalize the findings for other Transformer-based models, in the future. Worth mentioning is the implementation of the CDG task within the GraphCodeBERT framework, which was executed employing identical settings and hyperparameters as the other tasks and models. This uniform approach might contribute to the observed lower accuracy in the performance of the CDG task within the model.

Finally, this study was conducted before ChatGPT gone public. So a very relevant extension of this work would be including GPT-4 model both as a code model as well as a XAI method to provide explanations on the decisions of itself and other models.

\section{Conclusion and future works}
\label{Conclusion}
This paper proposes an approach for explaining pre-trained code models (e.g., CodeBert and GraphCodeBert), using their internal end-to-end attention mechanism, as the XAI method. Unlike most XAI research, where the explanation is applied to high-accuracy models to make sure the results are trustworthy, we have used XAI on both high (to find out what the models learn) and low-accuracy scenarios (to see when they do not perform well). Our findings not only provide observations about what these state-of-the-art Transformer-based models learn in terms of token type categories and why they underperform in some scenarios, but also suggest actionable recommendations, such as using more subjective evaluation metrics for CDG task, giving token types as additional input to the model, and manually amplifying attention scores for specific token types. In the future, we plan to extend this work by examining other downstream tasks, models, and XAI methods. In addition, we also plan to work on pre-trained code models by implementing the suggested recommendations from our observations. Finally, we plan to use GPT-4 model to both access its results as code model and use it as an XAI technique to explain why a particular output was provided by the model.

\bibliographystyle{IEEEtran}
\bibliography{bibfile.bib}

\end{document}
% end of file template.tex